\documentclass[runningheads,a4paper]{aa}    
\usepackage{graphicx,url,twoopt,natbib}
\usepackage{multirow}
\usepackage[varg]{txfonts}           
\usepackage{hyperref}                
\usepackage{xcolor}
\usepackage{float}

\hypersetup{
	colorlinks=true,   
	urlcolor=blue,     
	linkcolor=blue     
}

\bibpunct{(}{)}{;}{a}{}{,}    
\graphicspath{ {figures/} }

\makeatletter
\newcommand{\bibnote}[2]{\global\@namedef{#1note}{#2}}
\newcommand{\biblink}[2]{\global\@namedef{#1link}{#2}}
\newcommand{\Alfven}[1]{Alfv\'en}

\newcommand{\rom}[1]{\romannumeral #1\relax}

\makeatother

\makeatletter
\protected\def\stonyslink{%
	\def\hyper@linkstart##1##2{}\let\hyper@linkend\@empty}
\newcommandtwoopt{\citeads}[3][][]{%
	\href{http://ui.adsabs.harvard.edu/abs/#3/abstract}%
	{\stonyslink \citealp[#1][#2]{#3}}
	\biblink{#3}{\href{http://ui.adsabs.harvard.edu/abs/#3/abstract}{ADS}}}
\newcommandtwoopt{\citepads}[3][][]{%
	\href{http://ui.adsabs.harvard.edu/abs/#3/abstract}%
	{\stonyslink \citep[#1][#2]{#3}}
	\biblink{#3}{\href{http://ui.adsabs.harvard.edu/abs/#3/abstract}{ADS}}}
\newcommandtwoopt{\citetads}[3][][]{%
	\href{http://ui.adsabs.harvard.edu/abs/#3/abstract}%
	{\stonyslink \citet[#1][#2]{#3}}
	\biblink{#3}{\href{http://ui.adsabs.harvard.edu/abs/#3/abstract}{ADS}}}
\newcommandtwoopt{\citeyearads}[3][][]{%
	\href{http://ui.adsabs.harvard.edu/abs/#3/abstract}%
	{\stonyslink \citeyear[#1][#2]{#3}}
	\biblink{#3}{\href{http://ui.adsabs.harvard.edu/abs/#3/abstract}{ADS}}}
\makeatother

\begin{document}  
	
	
	\title{Validity of using Els\"asser variables to study the interaction of compressible solar wind fluctuations with a coronal mass ejection}
	\titlerunning{Using Els\"asser variables to study the interaction of solar wind fluctuations with a CME}
	
	
	\author{Chaitanya Prasad Sishtla\inst{1} \and
		Jens Pomoell\inst{1} \and
		Norbert Magyar\inst{2} \and
		Emilia Kilpua\inst{1} \and
		Simon Good\inst{1}
	}
	
	\institute{Department of Physics, University of Helsinki, PO Box 64, 00014 Helsinki, Finland\\
		\email{chaitanya.sishtla@helsinki.fi}
		\and
		Centre for mathematical Plasma Astrophysics, Mathematics Department, KU Leuven, Celestijnenlaan 200B bus 2400, B-3001 Leuven, Belgium
	}
	
	\date{}
	
	
	\abstract
	{Alfv\'enic fluctuations, as modelled by the non-linear interactions of Alfv\'en waves of various scales, are seen to dominate solar wind turbulence. However, there is also a non-negligible component of non-Alfv\'enic fluctuations. The Els\"asser formalism, which is central to the study of Alfv\'enic turbulence due to its ability to differentiate between parallel and anti-parallel Alfv\'en waves, cannot strictly separate wavemodes in the presence of compressive magnetoacoustic waves. In this study, we analyse the deviations generated in the Els\"asser formalism as density fluctuations are naturally generated through the propagation of a linearly polarised Alfv\'en wave. The study was performed in the context of a coronal mass ejection (CME) propagating through the solar wind, which enables the creation of two solar wind regimes, pristine wind and a shocked CME sheath, where the Els\"asser formalism can be evaluated.}
	{We studied the deviations of the Els\"asser formalism in separating parallel and anti-parallel components of Alfv\'enic solar wind perturbations generated by small-amplitude density fluctuations. Subsequently, we evaluated how the deviations cause a misinterpretation of the composition of waves through the parameters of cross helicity and reflection coefficient.}
	{We used an ideal 2.5D magnetohydrodynamic (MHD) model with an adiabatic equation of state. An Alfv\'en pump wave was injected into the quiet solar wind by perturbing the transverse magnetic field and velocity components. This wave subsequently generates density fluctuations through the ponderomotive force. A CME was injected by inserting a flux-rope modelled as a magnetic island into the quasi-steady solar wind.}
	{The presence of density perturbations creates a $\approx 10\%$ deviation in the Els\"asser variables and reflection coefficient for the Alfv\'en waves as well as a deviation of $\approx 0.1$ in the cross helicity in regions containing both parallel and anti-parallel fluctuations.}
	{}
	
	
	\maketitle
	%
	\section{Introduction}     \label{sec:introduction}
	The origin and acceleration of the solar wind can be explained through the turbulent cascade of large-wavelength Alfv\'enic perturbations to kinetic scales that heat the plasma~\citep{coleman1968turbulence, belcher1971large, goldreich1995toward, alazraki1971solar}. These Alfv\'en waves are generated through photospheric convective motions where the solar magnetic field footpoints are anchored~\citep{cranmer2005generation}. The presence of Alfv\'enic fluctuations has been observed both in situ~\citep{belcher1971large, dAmicis2015} and remotely~\citep{tomczyk2007alfven}, and their amplitudes have been estimated to be between $20~$km s$^{-1}$ and $55~$km s$^{-1}$ based on observations of non-thermal velocity amplitudes in the upper transition region of the Sun~\citep{chae1998sumer, doyle1998coronal}. Here, `Alfv\'enic fluctuations' refers to perturbations polarised perpendicular to the mean magnetic field $\mathbf{B}_0$ that exhibit correlations between the velocity and magnetic field components. However, solar wind fluctuations also have a measurable fraction of non-Alfv\'enic (compressible) modes~\citep{higdon1984density}. These fluctuations can arise through non-linear wave-wave interactions of the fast, slow, and Alfv\'en modes~\citep{nakariakov1997alfven, chandran2005weak, fu2022nature}, from instabilities~\citep{goldstein1978instability, Derby1978}, and from turbulence driving processes. Subsequently, in situ solar wind measurements beyond the Alfv\'en critical point have indicated that most fluctuation power is contained in the Alfv\'enic modes as opposed to the compressive fluctuations~\citep{bruno2013solar, chen2016recent}. Under the assumptions that the turbulent fluctuations are small compared to the mean field, are spatially anisotropic with respect to it, and have a frequency that is low compared to the ion cyclotron frequency, \citet{schekochihin2009astrophysical} showed that the `inertial range' cascade separates into two parts: a cascade of Alfv\'enic fluctuations and a passive cascade of density fluctuations. This allowed us to study the cascade of Alfv\'enic fluctuations in the incompressible limit with Alfv\'enic turbulence passively interacting with the compressive fluctuations. Thus, coronal heating via Alfv\'enic turbulence in global simulations~\citep{mikic1999magnetohydrodynamic, mikic2018predicting} has been modelled in the incompressible regime through the reflection-driven turbulence model~\citep{chandran2019reflection}. This model is supported by numerous studies covering various aspects of the reflection-driven turbulence: the linear Alfv\'en wave problem~\citep{Ferraro1958, goldreich1995toward, velli1993propagation, suzuki2004coronal, sishtla2022flux}; radial evolution of turbulence~\citep{verdini2007alfven, tenerani2017evolving, zank2018theory}; and incorporation of the turbulence model into simulations~\citep{cranmer2007self, chandran2011incorporating, van2014alfven}. Such turbulence models approach the heating problem by considering counter-propagating Alfv\'en waves generated through reflections from large-scale gradients in the solar wind~\citep{velli1989turbulent, zhou1989non}, and they model the turbulent heating due to incompressible fluctuations that have been found to dominate the solar wind in the heliosphere~\citep{tu1995mhd}. 
	
	An essential tool to analyse incompressible magnetohydrodynamic (MHD) turbulence is Els\"asser ~\citep{elsasser1950hydromagnetic} formalism, which in the solar wind enables the separation of sunward and anti-sunward directed Alfv\'enic fluctuations based on the correlation and anti-correlation of the velocity and magnetic field. The separation of sunward and anti-sunward directed Alfv\'enic fluctuations enables the modelling of coronal heating via Alfv\'enic turbulence through the interaction of counter-propagating Alfv\'en waves. The significance of the Els\"asser variables becomes less precise in the presence of density perturbations caused by magnetoacoustic waves. This is particularly evident when distinguishing between sunward and anti-sunward directed fluctuations~\citep{magyar2019nature}. Previously, \citet{marsch1987ideal} has shown that the compressible MHD equations can be expressed through Els\"asser variables with a variable density, with small-amplitude density perturbations allowing for decomposition of the variables into a purely Alfv\'enic and a compressive component. Therefore, Els\"asser variables need not preserve their decomposition of the waves propagating in opposite directions in the compressive MHD regime. However, the usage of these variables based on their ability to separate the directionality of waves is central in the reflection-driven turbulence model and widely used in the in situ analysis of the solar wind, particularly when it can be assumed that most of the wave power lies within the Alfv\'enic modes~\citep{tu1989basic, grappin1990origin, good2022cross}. However, the effect of small-amplitude density fluctuations on Alfv\'en wave dynamics is essential to enable wave reflections sufficient to accelerate the solar wind via turbulent heating~\citep{van2016heating}. Thus, it is imperative to consider Alfv\'en waves mixed with density fluctuations when analysing the plasma in simulations and observational data, especially as it is not possible to exactly decompose the waves into Alfv\'en and non-Alfv\'en waves due to their non-linear mixing~\citep{gan2022existence, fu2022nature}. Therefore, estimating the deviations in the Els\"asser formalism that are introduced by the density fluctuations is important due to the non-negligible component of compressible turbulence throughout the heliosphere~\citep{marsch1990spectral} and their importance in heating and accelerating the wind.
	
	In this study, we investigate the impact of compressive density fluctuations on the Els\"asser-based interpretation of the waves in the solar corona in the context of a coronal mass ejection (CME) interacting with solar wind fluctuations using MHD simulations. CMEs are transient eruptions of plasma and the magnetic field from the solar corona, and they often exhibit a three-part structure in coronagraph images consisting of a bright front of compressed coronal plasma enclosing a dark, low-density cavity (assumed to correspond to a magnetic flux rope, or FR) that contains a high-density
	core~\citep{gibson2000three, kilpua2017coronal}. In this study, we employed a simulation methodology similar to the simulation described in \citet{sishtla2023interact} (hereafter referred to as S23), which studies the interaction of a CME with (shear) Alfv\'en waves in the low corona. We also self-consistently generated compressive fluctuations due to the evolution of a shear Alfv\'en wave continually injected at the low coronal boundary. This contrasts with the simulation reported in S23, which contains only incompressible fluctuations. In S23, a lower simulation grid resolution causes the Alfv\'en waves to be damped due to numerical diffusion before the density fluctuations can be generated. The density fluctuations in the simulation of this work were generated through a ponderomotive force created by the propagating Alfv\'en wave (Section~\ref{sec:injectAW}). The presence of the CME is important to separating the solar wind plasma into two regimes: pristine wind upstream of the CME and a shocked CME sheath. This separation allows for an analysis of the fluctuations in the two regimes that exhibit different dynamics of the propagating wave due to the differing Alfv\'en and sound speeds. Upon restricting our analysis to frequencies close to the Alfv\'en wave frequency in the quiet wind, we found the shocked CME sheath structure in the simulation to exhibit minimal density fluctuations. This occurs as the shock compresses the upstream plasma, causing the CME sheath waves to propagate outside the frequency range we investigate. Thus, the shock appears to reset the solar wind fluctuations for the range of frequencies we study. 
	The separation of the pristine wind from the CME sheath allowed us to study the Els\"asser formalism in two regimes defined by density fluctuations through the same simulation. 
	
	The study finds the large-scale structures of the CME to be similar to those found in S23 but with the addition of density fluctuations in the simulation domain. However, we find that the small-amplitude compressive waves contribute to significant misinterpretations when analysing the Alfv\'enic waves using Els\"asser variables. We find that Els\"asser variables do not strictly allow for the separation between sunward and anti-sunward Alfv\'en waves in the presence of density fluctuations. The density fluctuations create a deviation of $\approx 10\%$ in the Els\"asser variables and the calculated reflection coefficient for the Alfv\'en waves and a deviation of $\approx 0.1$ in the cross helicity in regions containing balanced fluctuations. In Section~\ref{sec:method}, we introduce the MHD equations and associated boundary conditions, the mechanism for Alfv\'en wave injection, and the CME model used in the simulations. A discussion of the density fluctuations affecting the reflection of Alfv\'en waves due to Alfv\'en velocity gradients is presented in Section~\ref{sec:injectAW}. The validity of the Els\"asser variable formulation is discussed in Section~\ref{sec:validity-elsasser}, and the deviations introduced in the cross helicity and reflection coefficient through the use of the formalism are discussed in Section~\ref{sec:discussion}. Finally, the conclusions are summarised in Section~\ref{sec:conclusion}.
	
	\section{Methodology}     \label{sec:method}
	We performed a 2.5D MHD numerical simulation assuming a radially outward magnetic field to initialise the solar wind. The MHD equations were advanced in time using the strong stability preserving (SSP) Runge-Kutta method to advance the semi-discretised equations~\citep{pomoell2012influence}. The numerical method employed the Harten–Lax–van Leer (HLL) approximate Riemann solver supplied by piece-wise, linear slope-limited interface states. The equations were solved in spherical coordinates, and the magnetic field was ensured to be divergence free to the floating point accuracy by utilising the constrained transport method~\citep{kissmann2012semidiscrete}.
	
	\begin{figure}[ht]
		\centering
		\includegraphics[width=0.5\textwidth]{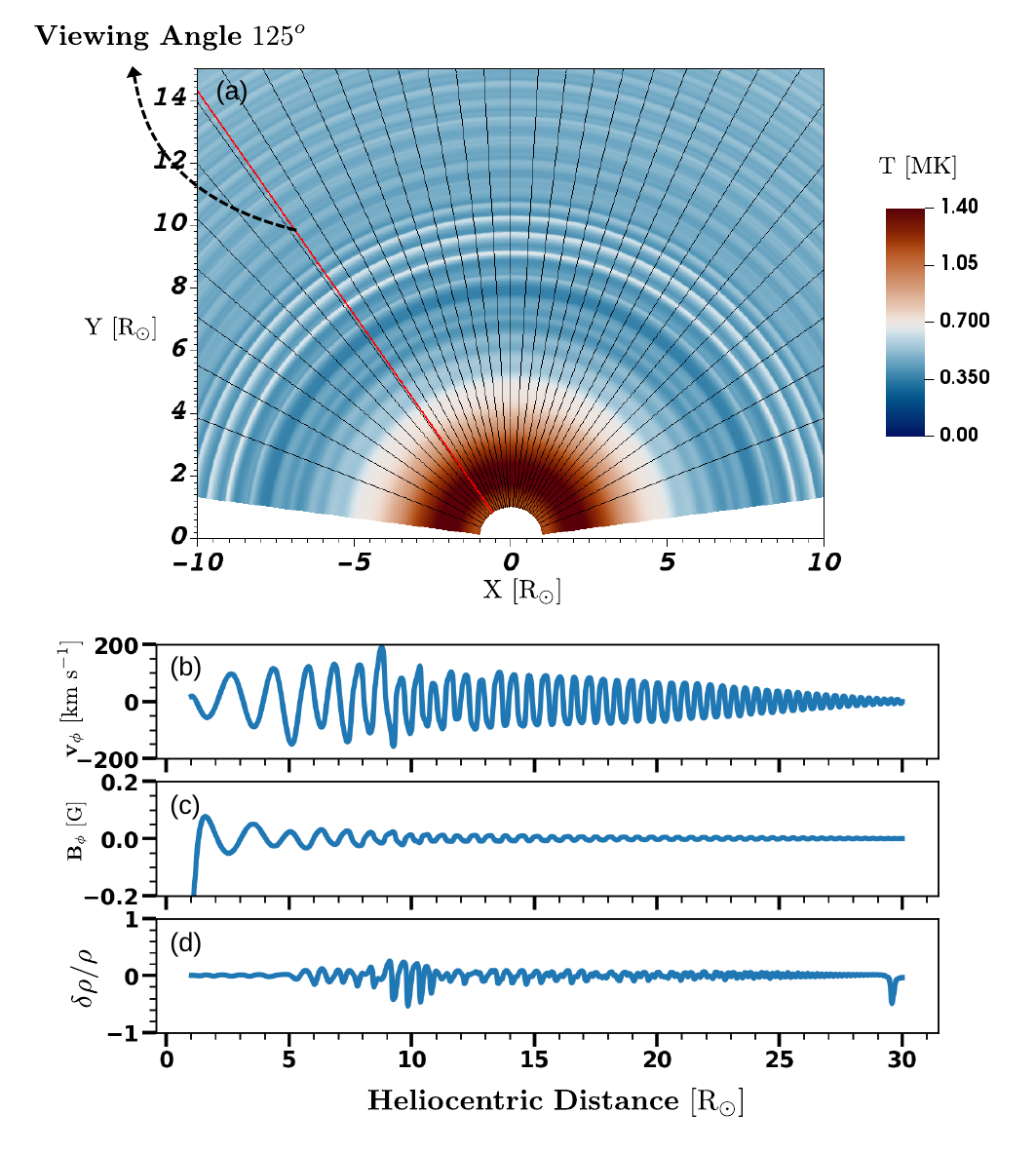}
		\caption{Coronal quasi-steady state. Panel (a) shows a simulation snapshot of the plasma temperature upon the injection of a $1~$mHz linearly polarised Alfv\'en wave, with an annotation describing the viewing angle along $125^\circ$ at $\tau = 1000$ mins after the injection of the Alfv\'en wave. Panels (b) and (c) show the out-of-plane $\mathbf{v_\phi}$ velocity and the $\mathbf{B_\phi}$ magnetic field components, respectively. The fluctuations induced in the density $\rho$ from the quasi-steady values prior to the injection of the Alfv\'en wave are presented in panel (d).}
		\label{fig:fig1}
	\end{figure}
	
	\begin{figure}[ht]
		\centering
		\includegraphics[width=0.5\textwidth]{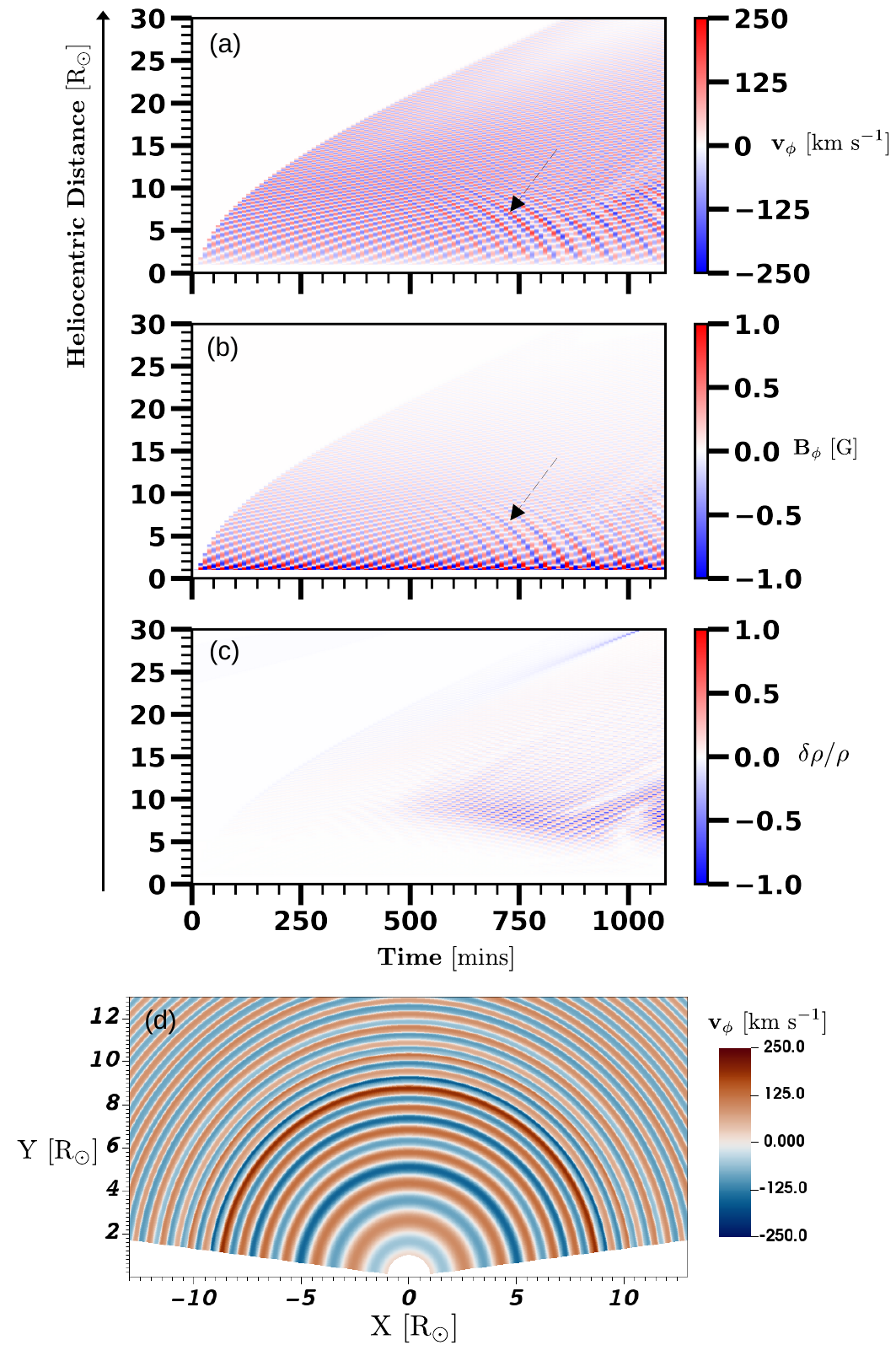}
		\caption{Evolution of the solar wind as the Alfv\'en waves propagate through the corona. The fluctuating velocity and magnetic field components are shown in panels (a) and (b). The density perturbations defined as $\delta\rho/\rho$, where $\delta\rho = \rho - <\rho>$ and 
			$<\rho>$ is the density averaged over $10$ min intervals, are shown in panel (c). Panel (d) is a simulation snapshot of $\mathbf{v_\phi}$ at $\tau=1000~$mins. The dashed arrows in panels (a) and (b) highlight the apparent bifurcation of $v_\phi$ and $B_\phi$.
		}
		\label{fig:BC_density}
	\end{figure}
	
	The MHD equations were integrated forward in time and in two spatial dimensions by considering a meridional plane with a radial extent of $r = 1.03~R_\odot$ to $r = 30~R_\odot$ and a co-latitudinal extent of $\theta = 10^\circ$ to $\theta = 170^\circ$. The simulation domain, therefore, exhibits rotational invariance in the out-of-plane $\phi$ direction. However, the vector quantities in the MHD equations retain all three components. The magnetic field was initialised to be purely radial and directed outwards and defined by the vector potential 
	$\mathbf{A} = -B_0 r_0 (r_0/r)\cot{\theta}~\hat{\mathbf{\phi}}$, where $B_0 = 5~G$ is the field strength at $r=r_0$, and the magnetic field was then specified using $\mathbf{B = \nabla}\times \mathbf{A}$. At the inner radial boundary, the density and temperature were chosen to be independent of latitude with $\rho_0 = 8.5\times 10^{-13}~\mathrm{kg}$ and $T_0 = 1.2\times 10^6~\mathrm{K}$. We linearly extrapolated all dynamical quantities in order to enforce an outflow boundary condition at the outer radial and latitudinal boundaries. The simulation domain was defined by a non-regular grid with $1763$ cells spaced logarithmically until $2.5~R_\odot$ and equidistantly spaced after that in the radial direction. In comparison, the simulation grid in S23 was defined by $500$ cells spaced logarithmically in the radial direction. The grid contained $128$ cells in the latitudinal direction both in the present simulation and in S23.
	
	The MHD equations with the relevant physical processes of gravity and ad hoc heating that are numerically solved are given in S23. The solar wind plasma evolves by solving these MHD equations for a polytropic index of $\gamma = 5/3$. An ideal gas law specified as $P = (\rho/m)k_B T$, where $m$ is the mean molecular mass, $k_B$ is the Boltzmann constant, and $P$ the pressure is used to compute the temperature $T$. We incorporated an additional energy source term to obtain a steady-state solar wind that approximates a Parker-like outflow~\citep{pomoell2015modelling, mikic2018predicting}.
	
	
	\subsection{Introducing Alfv\'enic perturbations}     \label{sec:injectAW}
	Once a steady-state solar wind was achieved after advancing the MHD equations in time, we introduced linearly polarised shear Alfv\'en waves by perturbing the $v_\phi$ and $B_\phi$ components at the low coronal boundary in the same manner as detailed in S23. The response of the solar wind to the introduction of the linearly polarised Alfv\'en wave is shown in Figure~\ref{fig:fig1}. Panel (a) presents a simulation snapshot of the temperature ($T$) at $\sim 16.6~$hours after the start of the injection of the Alfv\'en wave. The temperature increases from $1.2~$MK at the lower boundary to $1.4~$MK at $\approx 2.5~R_\odot$ before decreasing. We observed compressive waves throughout the simulation, as represented by the wave-like features in panel (a). To further illustrate the Alfv\'en and compressive waves, we included a plot of the $v_\phi$, $B_\phi$, and the density fluctuation level in panels (b), (c), and (d), respectively, along a radial ray at a viewing angle of $125^\circ$, as annotated in panel (a). To describe the propagation of density fluctuations, we defined the fluctuation in density $\delta\rho$ as $\rho = \rho_0 + \delta\rho$, where $\rho_0 (\tau)$ is the $10$-min time-averaged density. This time averaging of the density allowed us to capture fluctuations in $\delta\rho$ up to $\approx 1.6~$mHz. Panels (b) and (c) illustrate the presence of Alfv\'en wave modes in the simulation with the $\phi$ components of the magnetic field and velocity fluctuating in correlation. The initially injected Alfv\'en wave appears to steepen between $5-15~R_\odot$ before dissipating (panel (b)), with the fluctuations positively or negatively correlated with the fluctuations in $B_\phi$ (panel (c)). Finally, in panel (d), we observed density fluctuations throughout the simulation domain, albeit at varying levels.
	
	
	The dynamics of the propagating Alfv\'en wave in a homogeneous medium (in the co-latitudinal direction), such as in this simulation, depends on the wave polarisation. It has been shown that circularly polarised waves of arbitrary amplitudes are exact solutions of the MHD equations and exhibit no net magnetic field pressure variations as they propagate~\citep{Ferraro1958, goldstein1978instability}. 
	In contrast, linearly polarised Alfv\'en waves create magnetic pressure variations as they propagate, causing them to steepen~\citep{cohen1974nonlinear}. The magnetic pressure gradients are balanced by an oscillating thermal pressure. Subsequently, the oscillating thermal pressure generates a ponderomotive force, which creates density fluctuations in the compressible MHD regime~\citep{hollweg1971density,nakariakov1997alfven}.
	This process of steepening (causing a temperature increase) and generating compressive waves is described through Figure~\ref{fig:BC_density}. In panels (a), (b), and (c) of the figure, we present $v_\phi$, $B_\phi$, and $\delta\rho/\rho$ along a viewing angle of $125^\circ$ from the moment of Alfv\'en wave injection (defined here as $\tau = 0$) until the end of the simulation ($\tau = 1000$ min). The injected waves only reach the end of the simulation domain ($30~R_\odot$) at $\tau \approx 850~$mins. However, until $\tau\approx 500~$mins, there are minimal density fluctuations as the simulation evolves towards a quasi-steady state. Then, from $\tau\approx 500~$mins to $\tau\approx 1000~$mins, the injected Alfv\'en wave grows linearly as the wave steepens (Figure~\ref{fig:fig1}(a) plotted at $\tau = 1000~$mins), and there is continual generation of compressive wave modes. In Figure~\ref{fig:BC_density}(d), we present a simulation snapshot of ${v_\phi}$ at $\tau = 1000~$mins depicting the homogeneous evolution of the injected Alfv\'en wave, as the crests and troughs of the wave do not have an angular dependence. 
	
	\begin{figure}[ht]
		\centering
		\includegraphics[width=0.5\textwidth]{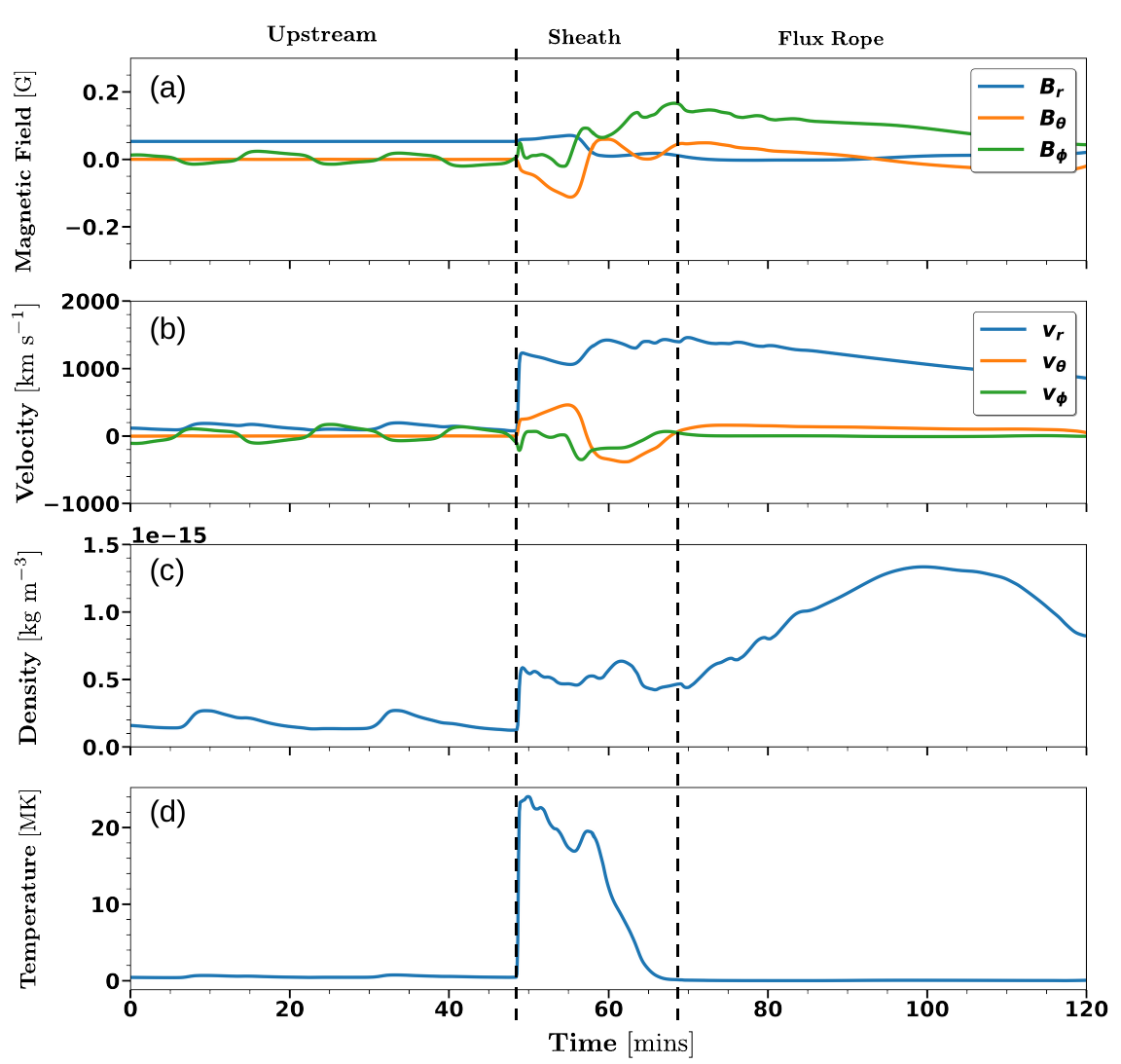}
		\caption{Measurement of plasma parameters by a virtual spacecraft located at 10 $R_\odot$ and with a viewing angle of $125^\circ$. The vertical lines delineate the upstream, sheath, and FR intervals encountered by the spacecraft.}
		\label{fig:virtual-craft}
	\end{figure}
	
	In Figure~\ref{fig:BC_density}, we observed that the Alfv\'en waves injected at the lower boundary experience reflections at distances of $10-15~R_\odot$. This reflection is evident in the apparent bifurcation of the $v_\phi$ component (panel (a)), which takes two different paths, as highlighted by the dashed arrow in panel (a). These paths correspond to the waves moving at their own sunward propagating phase speed $v - v_a$, where $v_a$ is the Alfv\'en speed, and a path corresponding to reflections from the anti-sunward wave and thus propagating with phase speed $v + v_a$~\citep{verdini2007alfven}. This region of $10-15~R_\odot$ corresponds to background inhomogeneities in the solar wind that are sufficient to generate Alfv\'en wave reflections. However, the density fluctuations generated by the ponderomotive force still propagate with the same velocity as the anti-sunward Alfv\'en wave that generates them. Subsequently, the density fluctuations lead to the formation of gradients in the Alfv\'en speed that are on a spatial scale comparable to the wavelength of the Alfv\'en wave. In summary, the excited anti-sunward wave generates density fluctuations via the ponderomotive force, and the anti-sunward wave is then reflected by the large-scale density gradients present in the solar wind. This reflected wave interacts with not only the anti-sunward wave but also with the density fluctuations that cause further scattering.
	
	Therefore, in this simulation, the Alfv\'en waves experience reflections from both large-scale background gradients as well as from small-scale variations in Alfv\'en velocity formed by the ponderomotive density fluctuations. In the absence of density fluctuations (as in S23), the Alfv\'en waves would only be reflected by background inhomogeneities.
	
	\subsection{Injecting coronal mass ejections}     \label{sec:injectCME}
	\begin{figure*}[ht]
		\centering
		\includegraphics[width=\textwidth]{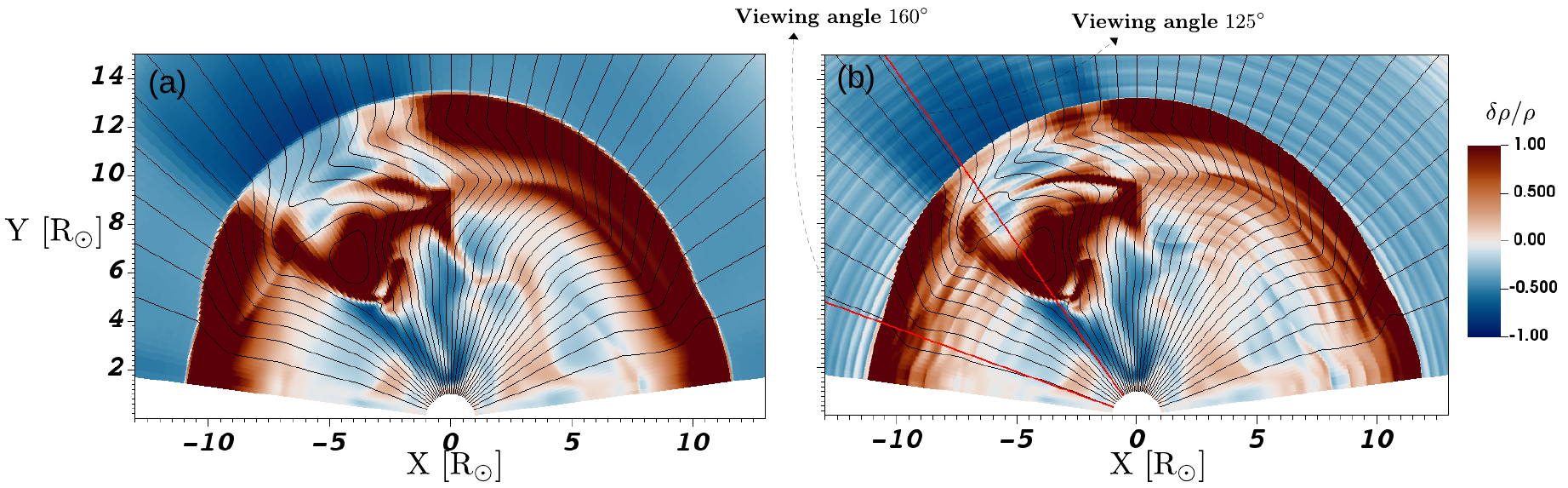}
		\caption{Simulation snapshots of density fluctuations at $t = 70$ min as the CME propagates through the solar wind. Panel (a) depicts the case from S23 where no compressible fluctuations exist, and panel (b) depicts the solar wind from the present study containing both Alfv\'enic and compressible fluctuations. Panel (b) is annotated with the viewing angles corresponding to the CME flank ($160^\circ$) and head-on ($125^\circ$).}
		\label{fig:paper-comp}
	\end{figure*}
	
	\begin{figure}[ht]
		\centering
		\includegraphics[width=0.48\textwidth]{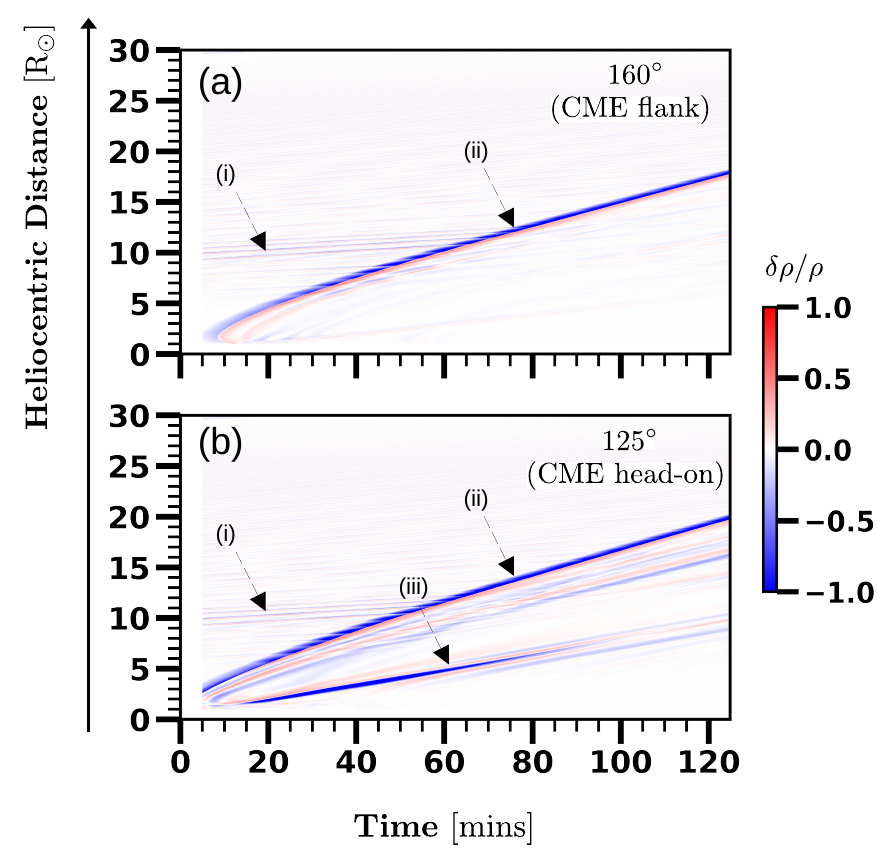}
		\caption{Ten-minute averaged density perturbations $\delta\rho/\rho$ are presented for viewing angles of $160^\circ$ (flank, panel (a)), and $125^\circ$ (head-on, panel (b)). The annotations denote enhanced perturbations near $10~R_\odot$ in the quiet wind (\rom{1}), the CME shock (\rom{2}), and the FR leading edge (\rom{3}).}
		\label{fig:results/density_waves}
	\end{figure}
	
	We chose the solar wind at $t=1000~$mins (Figure~\ref{fig:BC_density}) to represent the quasi-steady state in which we introduce the CME. We modelled the CME as a magnetic island with an FR magnetic field and a non-uniform density profile to populate the ejecta. The FR was modelled using the Soloviev solution to the Grad-Shafranov (GS) equation, which represents axisymmetric MHD equilibria of magnetised plasmas without flows such that the equilibrium condition $\mathbf{J}\times \mathbf{B} =\nabla P$ is satisfied. 
	A detailed explanation of the CME model and its eruption can be found in S23. We note that we do not model the CME eruption self-consistently. Rather, the specification of thermal pressure inside the CME and the superposition of the structure on the quasi-steady solar wind results in a non-equilibrium state, causing the FR to dynamically evolve by propagating and expanding.
	
	In Figure~\ref{fig:virtual-craft}, we present the solar wind parameters as a function of time as the CME is encountered by a virtual spacecraft located at $10~R_\odot$ and with a viewing angle of $125^\circ$. Before encountering the CME (i.e. in the upstream), the solar wind contains fluctuations in ${v_\phi}$ and ${B_\phi}$ (panels (a) and (b)) along with the density and temperature fluctuations (panels (c) and (d)) as the linearly polarised Alfv\'en wave generates compressive waves. The arrival of the CME is characterised by the CME-driven leading shock, as seen by the simultaneous jump in radial velocity, density, temperature, and magnetic field. Following the shock is the CME sheath, which is the compressed region of plasma driven by the propagating and expanding CME. We observed solar wind fluctuations, evident as variations in ${v_\phi}$ and ${B_\phi}$; they are present in the upstream and in the CME sheath. Also, a local enhancement in density at $t\approx 62$ mins corresponding to a pile-up compression region (PUC)~\citep{das2011evolution} is noticeable in the sheath. Finally, the spacecraft encounters the FR characterised by the smooth, large-scale rotation of the magnetic field's ${B_\theta}$ component. Hence, Figure~\ref{fig:virtual-craft} demonstrates the large-scale features typically observed in association with CMEs, namely, a shock, sheath, FR, and PUC.
	
	\begin{figure*}[ht]
		\centering
		\includegraphics[width=\textwidth]{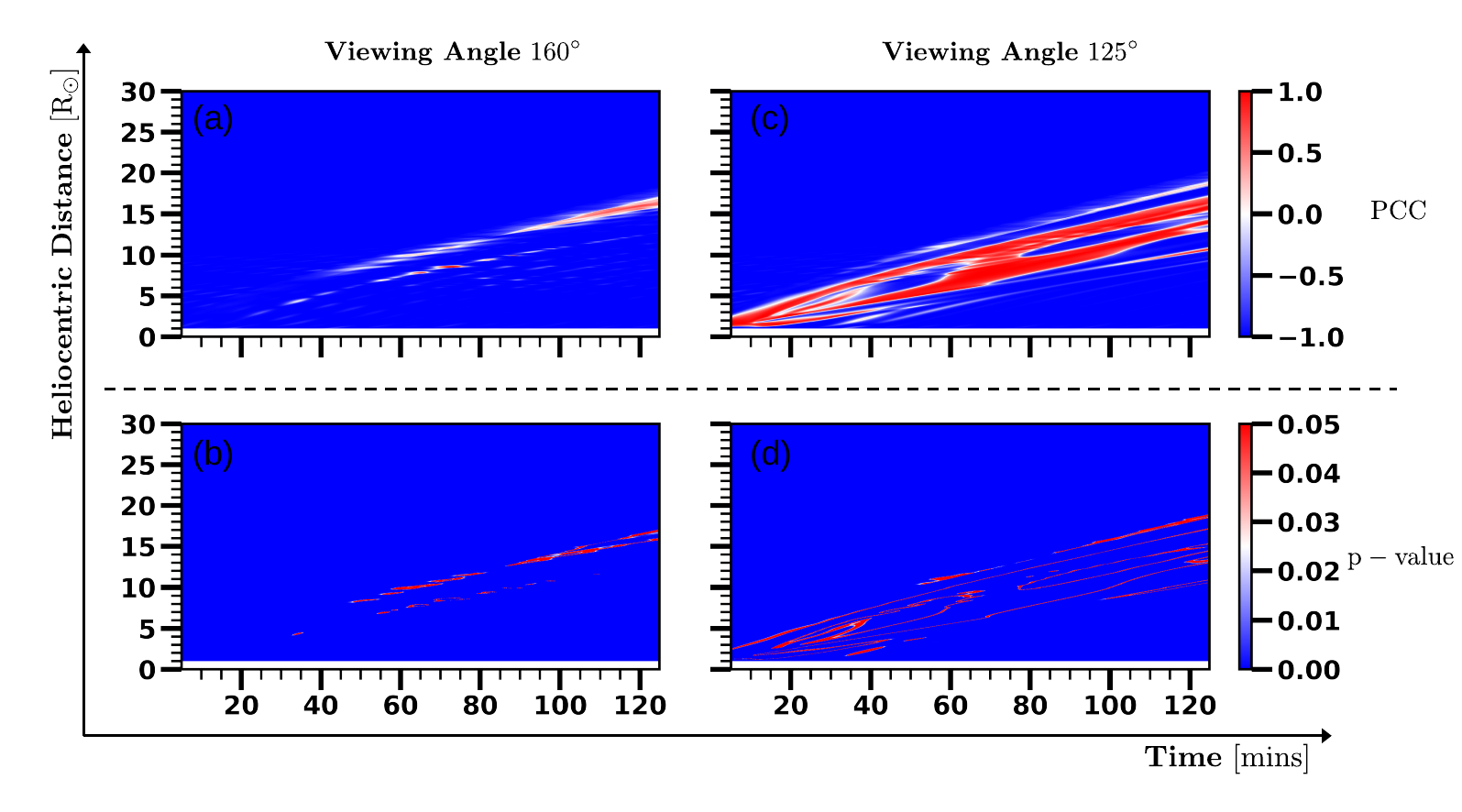}
		\caption{Pearson correlation coefficients (panels (a) and (c)) and the corresponding p-values (panels (b) and (d)) along viewing angles of $160^\circ$ (flank) and $125^\circ$ (head-on).}
		\label{fig:pearsonr}
	\end{figure*}
	
	\begin{figure*}[ht]
		\centering
		\includegraphics[width=\textwidth]{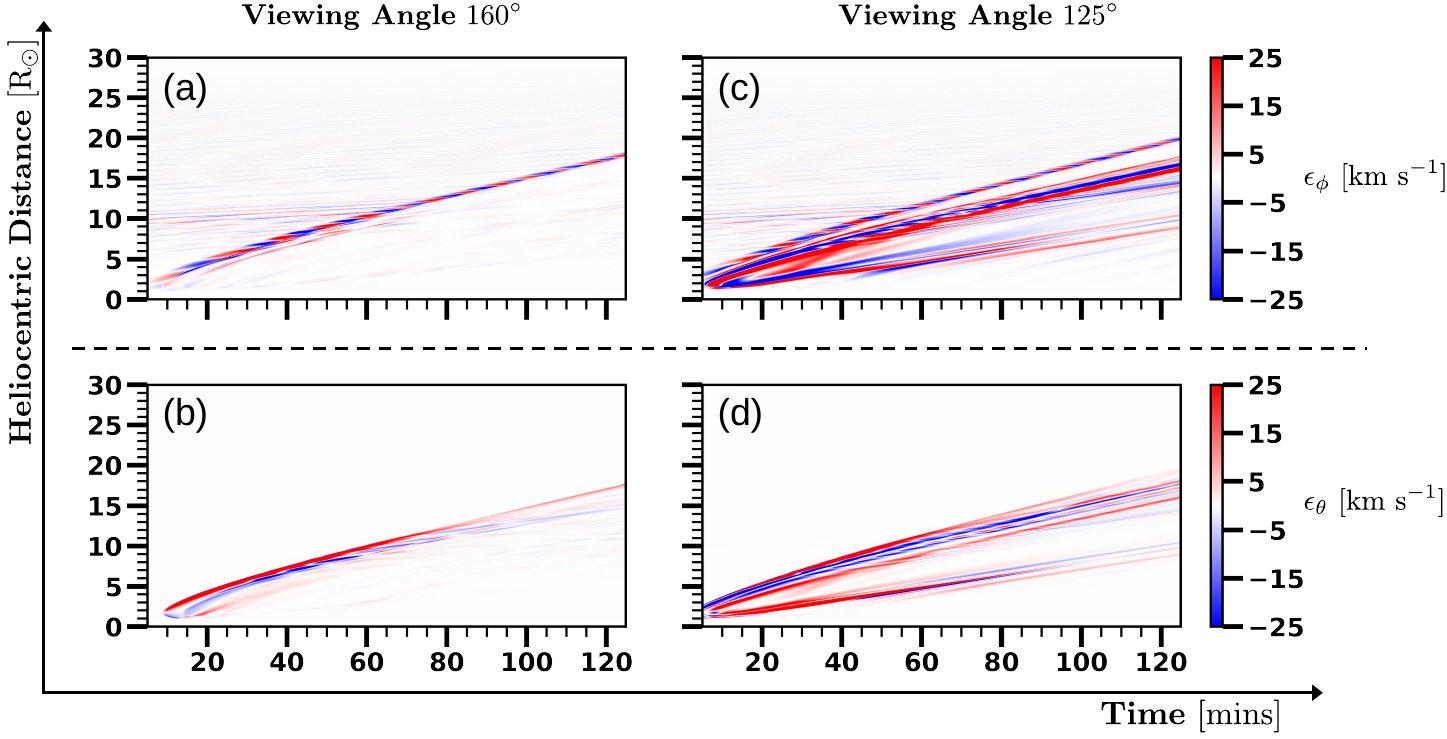}
		\caption{Deviations in the Els\"asser variables due to density fluctuations. The deviations ($\epsilon$) as defined in Equation~\ref{eq:elss-eps} are shown for the $\phi$ (panels a and c) and $\theta$ (panels b and d) directions along viewing angles of $160^\circ$ and $125^\circ$.}
		\label{fig:elssaser_error}
	\end{figure*}
	
	The modelling and analysis of a similar CME propagating in a solar wind containing only incompressible fluctuations is described in S23. In that study, the authors describe the transmission of upstream solar wind fluctuations into the CME sheath and the effect of the fluctuation frequency on the CME shock formation. The current simulation presents the evolution of the same CME in a solar wind containing compressible fluctuations. This distinction is illustrated in Figure~\ref{fig:paper-comp} with simulation snapshots of the density fluctuations ($\delta\rho/\rho$) at $t = 70~$mins. The snapshot in panel (a) shows the CME passage in a wind containing only incompressible waves. In panel (b), the solar wind contains density fluctuations as well, as described in Section~\ref{sec:injectAW}. The CME structure, on large scales, is  similar in both simulation runs, and the CME deflects in the same direction. The primary and stark difference between the two simulations is the density fluctuations, which are ubiquitous in both the pristine wind and in the CME sheath and clearly visible in panel (b) where compressible fluctuations are present.
	
	The presence of density perturbations and their interaction with the CME is presented in Figure~\ref{fig:results/density_waves}. Panels (a) and (b) show the perturbations in density $\delta\rho/\rho$ for viewing angles of $160^\circ$ (CME flank) and $125^\circ$ (CME head-on), respectively. In the pristine solar wind, we observed the density fluctuations in a manner consistent with Figure~\ref{fig:BC_density}(c), that is, as seen by the region of enhanced density perturbations near $10~R_\odot$ (annotated with (\rom{1})). When the density fluctuations encounter the CME-driven shock (annotated with (\rom{2})), their frequency increases as the shock compresses the plasma. Thus, we observed minimal perturbations of the $10$-min averaged density perturbations in the CME flanks (panel (a)) compared to the pristine solar wind. This indicates predominantly higher frequency density fluctuations as the upstream waves are compressed at the shock. Similarly, when viewing the CME head-on (panel b), we saw a region after the shock where the fluctuations propagate prior to encountering the FR leading edge (annotated with (\rom{3})). This can also be observed in Figure \ref{fig:paper-comp}(b), where one can observe density fluctuations along $125^\circ$ prior to the FR leading edge. The density fluctuations start recovering after the passage of the FR, as the Alfv\'en waves are continually injected at the lower boundary.
	
	\section{On the validity of the Els\"asser formalism} \label{sec:validity-elsasser}
	\begin{figure*}[ht]
		\centering
		\includegraphics[width=\textwidth]{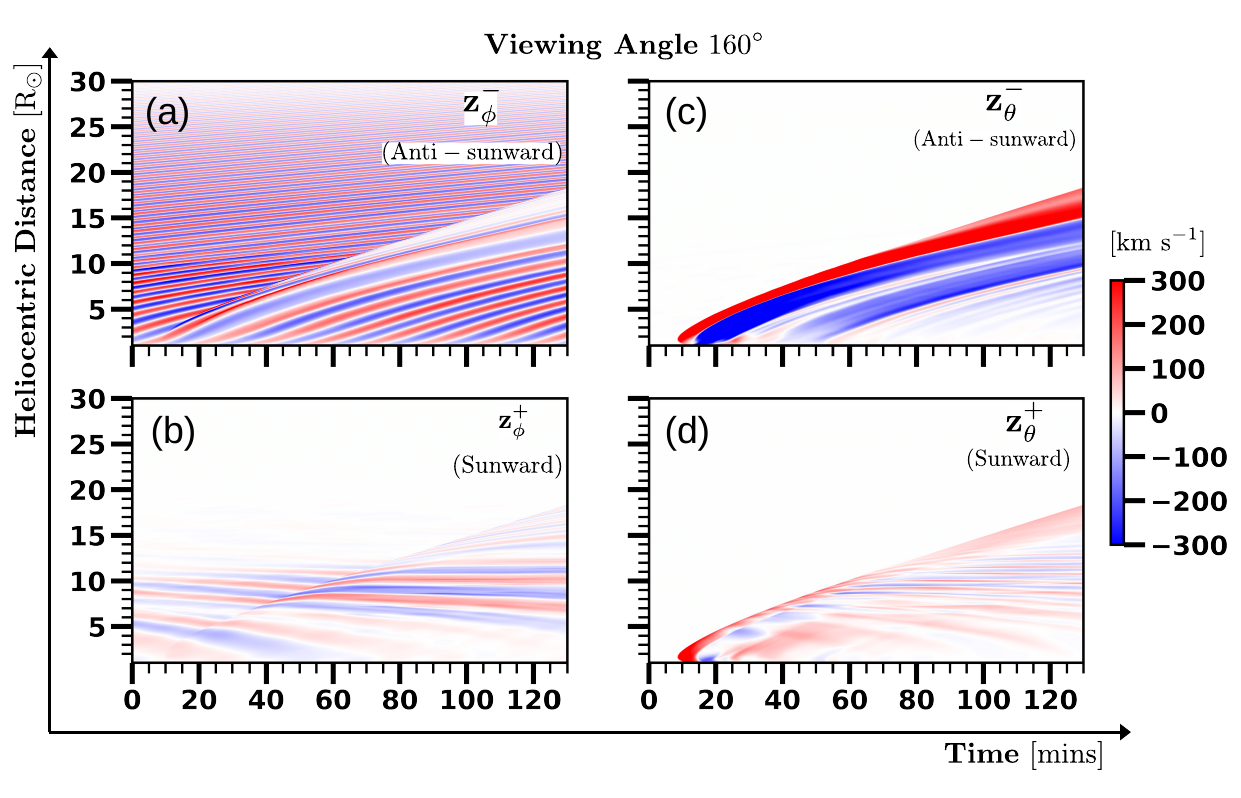}
		\caption{Els\"{a}sser variables along the $\phi$ and $\theta$ directions shown for a viewing angle of $160^\circ$ and an injected Alfv\'enic fluctuation frequency of $1$~mHz.}
		\label{fig:elssaser_160}
	\end{figure*}
	
	The simulation described in Section~\ref{sec:method} characterises a wind with primarily Alfv\'en waves and small-amplitude density fluctuations (without magnetosonic waves). As the properties of the solar wind fluctuations that have entered the CME sheath are primarily determined by their interaction with the CME shock (discussed in S23), the sheath similarly contains Alfv\'en waves mixed with density fluctuations. In this section, we discuss the limitations of Els\"asser variables in describing such a plasma.
	
	The Alfv\'enic nature (or Alfv\'enicity) of the waves is contained in the $\mathbf{v} - \mathbf{B}$ correlation that allows us to represent the waves as the fluctuating part of $\mathbf{z^\pm} = \mathbf{v} \pm  {\mathbf{B}}/{\sqrt{\mu_0\rho}}$. In an incompressible medium, plasma exhibits full Alfv\'enic behaviour, with all wave modes as shear Alfv\'en waves expressed using Els\"asser variables. Additionally, the $\mathbf{z^\pm}$ variables define wave directionality, that is, sunward or anti-sunward. 
	However, the density fluctuations in the upstream solar wind and downstream of the CME-driven shock first required us to verify the Alfv\'enicity of the waves. In Figure~\ref{fig:pearsonr}, we show the Pearson correlation coefficient (PCC) computed for a $10$-min averaging period and the corresponding p-value of the PCC along viewing angles of $160^\circ$ (CME flank) and $125^\circ$ (CME head-on). In the figure, the PCC is calculated between $v_\phi$ and $B_\phi$ and can denote anti-correlation (PCC $\in \, [-1, 0[$ with -1 indicating perfect anti-correlation), positive correlation (PCC $\in \, ]0, 1]$), or no correlation (PCC $=0$). The presence of correlation signifies the Alfv\'enic nature of the flows. The associated p-value is the probability of finding a correlation given that the parameters were uncorrelated initially (null hypothesis). Thus, in panels (b) and (d), the locations with a p-value larger than $0.05$ imply the probability of a false correlation as being greater than $5\%$.
	
	Figure~\ref{fig:pearsonr} shows a predominance of highly anti-correlated (anti-sunward propagating) waves except near the CME shock, which coincides with locations of enhanced density (Figure~\ref{fig:results/density_waves}). The associated p-values confirm the presence of non-Alfv\'enic and positively correlated waves in these regions. However, the PCC values cannot be trusted around the CME shock (panels (a) and (b)) and near the FR (panels (c) and (d)), as indicated by the greater than $0.05$ p-values. These regions of high p-values are locations where the plasma is quickly compressed either by the shock or the FR ejecta compressing the wind ahead of it.
	
	If we suppose the simulation had an instability, such as the parametric decay instability (PDI)~\citep{sagdeev1969nonlinear, Derby1978, goldstein1978instability, shoda2018frequency, chandran2018parametric, sishtla2022flux}, driving small-amplitude density fluctuations through the formation of MHD sound waves, we might still expect $\mathbf{v}-\mathbf{B}$ correlations, but the presence of the magnetosonic wave would lead to the wind not being purely Alfv\'enic. However, specifying how density fluctuations form in the present simulation (through the ponderomotive force) without the additional magnetosonic waves ensures that the observed $\mathbf{v}-\mathbf{B}$ correlation is in agreement with the presence of shear Alfv\'en waves, thus simplifying the definition of `Alfv\'enicity' in this context. Therefore, this confirmation of the Alfv\'enic nature of the waves supports our understanding that the solar wind fluctuations in our simulation combine shear Alfv\'en waves with density fluctuations. With this understanding confirmed, we could then investigate whether the directionality of the waves (sunward or anti-sunward propagating) is still captured by the definition of the Els\"asser variables. Since the imposed fluctuations are confined to the $\phi$ direction, we analysed the $\mathbf{z_\phi^\pm}$ component of the Els\"asser variables:
	\begin{align}
		\mathbf{z_\phi^\pm} = \mathbf{v_\phi} \pm \frac{\mathbf{B_\phi}}{\sqrt{\mu_0 (\rho_0 + \delta\rho)}}.
		\label{eq:full-def}
	\end{align}
	Without density fluctuations, the $\mathbf{z_\phi}^+$ and $\mathbf{z_\phi}^-$ variables would denote anti-sunward and sunward fluctuations, respectively. If we assume small-amplitude density fluctuations ($\delta \rho \ll \rho_0$), we can write
	\begin{align}
		\mathbf{z_\phi^\pm} = \underbrace{\mathbf{v_\phi} \pm \frac{\mathbf{B_\phi}}{\sqrt{\mu_0\rho_0}}}_{\mathbf{z}^\pm_{\phi,0}} \mp \underbrace{\frac{\delta\rho}{2\rho_0}\frac{\mathbf{B_\phi}}{\sqrt{\mu_0\rho_0}}}_{\Delta} \pm \ldots
		\label{eq:elss-error}
	\end{align}
	using a Taylor series expansion. Here, $\mathbf{z}^\pm_{\phi,0}$ indicates the Els\"asser variable in an incompressible medium and $\Delta$ is the leading order deviation due to the density fluctuations. Such an analysis has been performed by \citet{magyar2019nature}, and they showed that even in the absence of sunward fluctuations, the magnetoacoustic waves are necessarily described by both $\mathbf{z_\phi^\pm}$, as the $\mathbf{v}-\mathbf{B}$ correlations are not exact for non-Alfv\'en waves. Without making assumptions of the small-amplitude nature of the density fluctuations, we defined $\epsilon$ to quantify the difference between the Els\"asser variables with and without density fluctuations,
	\begin{align}
		\epsilon = \frac{\mathbf{B}}{\sqrt{\mu_0\rho}} - \frac{\mathbf{B}}{\sqrt{\mu_0\rho_0}}, 
		\label{eq:elss-eps}
	\end{align}
	where $\rho = \rho_0 + \delta\rho$. The $\epsilon$ parameter is calculated in Figure~\ref{fig:elssaser_error} for both the $\phi$ (panels (a) and (c)) and $\theta$ (panels b and d) directions along the viewing angles of $160^\circ$ (flank) and $125^\circ$ (head-on). In the upstream solar wind, we saw the presence of wave-like specks of $\epsilon_\phi$ (panels (a) and (c)), which appear to follow the pattern of density fluctuations in Figure~\ref{fig:results/density_waves}. The region of enhanced $\epsilon_\phi$ near $10~R_\odot$ coincides with the location of enhanced density fluctuations in Figure~\ref{fig:BC_density}(c) and Figure~\ref{fig:results/density_waves}. In panels (b) and (d), we observed no $\epsilon_\theta$ in the upstream wind since ${B_\theta} = 0$. After the CME shock transition (panels (a) and (b)), we observed negligible or zero $\epsilon_{\phi, \theta}$, as the plasma compression at the shock caused negligible $\delta\rho/\rho$ in the $10$-min averaging interval we considered. As the solar wind recovered, we started observing some $\epsilon_\phi$ at lower heliocentric distances that coincide with locations in Figure~\ref{fig:results/density_waves}(a) where density perturbations start propagating. A similar behaviour was observed when viewing the CME head-on (panels (c) and (d)), where we observed enhanced $\epsilon_{\phi, \theta}$ in locations of Alfv\'enic waves (Figure~\ref{fig:pearsonr}) containing compressed plasma (Figure~\ref{fig:results/density_waves}(b)) before encountering the FR. In Figure~\ref{fig:elssaser_error}, the $\epsilon$ magnitude is about $5-10$~km s$^{-1}$ in the pristine solar wind at locations with density fluctuations (such as the region annotated as (\rom{1}) in Figure~\ref{fig:results/density_waves}) and approximately $25$~km s$^{-1}$ close to the shock and CME sheath (near the region annotated as (\rom{2}) in Figure~\ref{fig:results/density_waves}). We note that the definition of the $\epsilon$ parameter in Equation~\ref{eq:elss-eps} depends on the averaging interval used to calculate $\rho_0$.
	
	The parameter $\epsilon_{\theta,\phi}$ indicates that even in the absence of either sunward or anti-sunward waves, we would still see a minimum value of non-zero $\mathbf{z}^\pm_\phi$. To compare how the $5-10$~km s$^{-1}$ margin of deviation ($\epsilon$) interferes with our interpretation of directionality from the Els\"asser variables, we plotted the complete Els\"asser variables for the $\phi$ and $\theta$ directions at a viewing angle of $160^\circ$ in Figure~\ref{fig:elssaser_160}. Panel (a) depicts the transmission of the upstream anti-sunward Els\"asser variable into the CME sheath, leading to the formation of CME sheath fluctuations. The fluctuations in the CME sheath exhibit a high frequency (short wavelength) due to compression by the CME shock~\citep{vainio1998alfven, vainio1999self, sishtla2023interact}. The accompanying sunward component (panel (b)) is similarly transferred downstream, with the shock compression being less prominent for such waves. We direct readers to S23 for a detailed analysis of the transmission of upstream solar wind fluctuations into the CME sheath and their subsequent influence on the CME sheath's formation. 
	In the upstream wind, the $\mathbf{z}^+_\phi$ component is present in locations $< 10R_\odot$ as the anti-sunward wave is reflected from the large-scale density gradient. The Els\"asser variables in the $\theta$ direction (panels (c) and (d)) indicate the absence of fluctuations in the pristine wind, as the injected Alfv\'en wave was polarised in the $\phi$ direction. However, in the shock neighbourhood, the plasma experiences large flows, as seen by the enhanced $\mathbf{z^-_\theta}$ prior to the wave-like features observed after the passage of these large flows. These large flows are generated as a consequence of the non-radial shock formed by the draping of field lines around the FR. The $\mathbf{z}^+_\theta$ component shows fluctuations downstream of the shock with a comparable amplitude and propagation velocity as $\mathbf{z}^+_\phi$. We note that such wave-like features in $\mathbf{z}^\pm_\theta$ are absent when only incompressible shear Alfv\'en waves are present in the upstream wind (as in S23), indicating that they might be generated as a result of the scattering of $\phi$ polarised Alfv\'en waves by the density fluctuations.
	
	Focusing on the $\phi$ direction (Figure~\ref{fig:elssaser_160}(a),~(b)), we observed that the anti-sunward fluctuations have an amplitude of $\approx 100-250$~km s$^{-1}$, while the sunward components are $\approx 50$~km s$^{-1}$. Comparing with the regions of non-zero $\epsilon_\phi$ in Figure~\ref{fig:elssaser_error}(a), which have a $5-10$~km s$^{-1}$ margin of $\epsilon$, results in a $\approx 4\%$ (anti-sunward) and $\approx 10\%$ (sunward) deviation margin near $10~R_\odot$, where we observed enhanced density fluctuations. This maximum of the deviation margins is obtained through accounting for only density fluctuations with time periods of ten minutes, as dictated by the averaging interval. Therefore, shear Alfv\'en waves in a solar wind plasma containing density fluctuations at a level of $\delta\rho/\rho \approx 0.1-0.25$ cannot exactly be decomposed into sunward and anti-sunward components. The deviations become more pronounced as the density perturbations increase in amplitude (Equation~\ref{eq:elss-error}). For the case of  density fluctuations at smaller amplitude levels, such as in this simulation, one can still observe a non-negligible difference of $\approx 4\%-10\%$ in the overall amplitude of the Alfv\'enic waves.
	
	\section{Composition of the Alfv\'enic fluctuations}     \label{sec:discussion}
	A significant utility of Els\"asser variables in the analysis of incompressible fluctuations is revealing the composition of waves in the plasma, for example, through the cross helicity and reflection coefficient. With the assumption of an incompressible plasma, cross helicity is a rugged invariant~\citep{matthaeus1982measurement} and can therefore track the evolution of sunward and anti-sunward Alfv\'en waves in the plasma. In this section, we investigate the misinterpretations caused by the cross helicity and reflection coefficient measures due to the deviations in the Els\"asser formalism generated by the presence of density fluctuations.
	
	\subsection{cross helicity}
	The cross helicity parameter ($\sigma_c$) describes the alignment between the Els\"asser variables ($\mathbf{z^\pm}$) and measures the difference in power between the counter-propagating Alfv\'enic fluctuations. It is defined as
	\begin{align}
		\sigma_{c} = \frac{|\mathbf{z^+}|^2 - |\mathbf{z^-}|^2}{|\mathbf{z^+}|^2 + |\mathbf{z^-}|^2}.
	\end{align}
	For simplicity, if we restrict ourselves to the $\phi$ direction,
	\begin{align}
		\sigma_{c,\phi} = \frac{|\mathbf{z^+_\phi}|^2 - |\mathbf{z^-_\phi}|^2}{|\mathbf{z^+_\phi}|^2 + |\mathbf{z^-_\phi}|^2} = \frac{{z^+_\phi}^2 - {z^-_\phi}^2}{{z^+_\phi}^2 + {z^-_\phi}^2}.
		\label{eq:ch-def}
	\end{align} 
	We can simplify $\sigma_{c,\phi}$ further using Equation~\ref{eq:elss-error} to obtain
	\begin{align}
		\sigma_{c,\phi} = \frac{({z}^+_{\phi,0} - |\Delta|)^2 - ({z}^-_{\phi,0} + |\Delta|)^2}{({z}^+_{\phi,0} - |\Delta|)^2 + ({z}^-_{\phi,0} + |\Delta|)^2}
		\\ = \frac{\{{{z}^+_{\phi,0}}^2 + |\Delta|^2 - 2{z}^+_{\phi,0}|\Delta|\} - \{{{z}^-_{\phi,0}}^2 + |\Delta|^2 + 2{z}^-_{\phi,0}|\Delta|\}}{\{{{z}^+_{\phi,0}}^2 + |\Delta|^2 - 2{z}^+_{\phi,0}|\Delta|\} + \{{{z}^-_{\phi,0}}^2 + |\Delta|^2 + 2{z}^-_{\phi,0}|\Delta|\}}.
	\end{align}
	By writing the result in the form
	\begin{align}
		\sigma_{c,\phi} = \frac{({{z}^+_{\phi,0}}^2 - {{z}^-_{\phi,0}}^2) - 2|\Delta|({z}^+_{\phi,0} + {z}^-_{\phi,0})}{({{z}^+_{\phi,0}}^2 + {{z}^-_{\phi,0}}^2) - 2|\Delta|({z}^+_{\phi,0} - {z}^-_{\phi,0})}
		\label{eq:sigma-expansion},
	\end{align}
	a non-trivial dependence of the cross helicity on $\delta\rho$ becomes evident. If only sunward ($\mathbf{z}^-_{\phi,0} = 0$) or anti-sunward ($\mathbf{z}^+_{\phi,0} = 0$) waves are present, then Equation~\ref{eq:sigma-expansion} still reduces to 1 or -1, respectively. Instead, if 
	the system exhibits a balanced distribution of waves (i.e. $|\mathbf{z}^-_{\phi,0}| = |\mathbf{z}^+_{\phi,0}| = {z}_{\phi,0}$), then Equation~\ref{eq:sigma-expansion} reduces to $-|\Delta|/{z}_{\phi,0}$. Assuming a solar wind similar to Figures~\ref{fig:elssaser_error} and \ref{fig:elssaser_160} with $|\Delta| \approx \epsilon_\phi \approx 15~$km s$^{-1}$ and ${z}_{\phi,0} \approx 100~$km s$^{-1}$, we can calculate $-|\Delta|/{z}_{\phi,0} = -0.15$, which would be a noticeable deviation from the expected zero value.
	
	\begin{figure*}[ht]
		\centering
		\includegraphics[width=\textwidth]{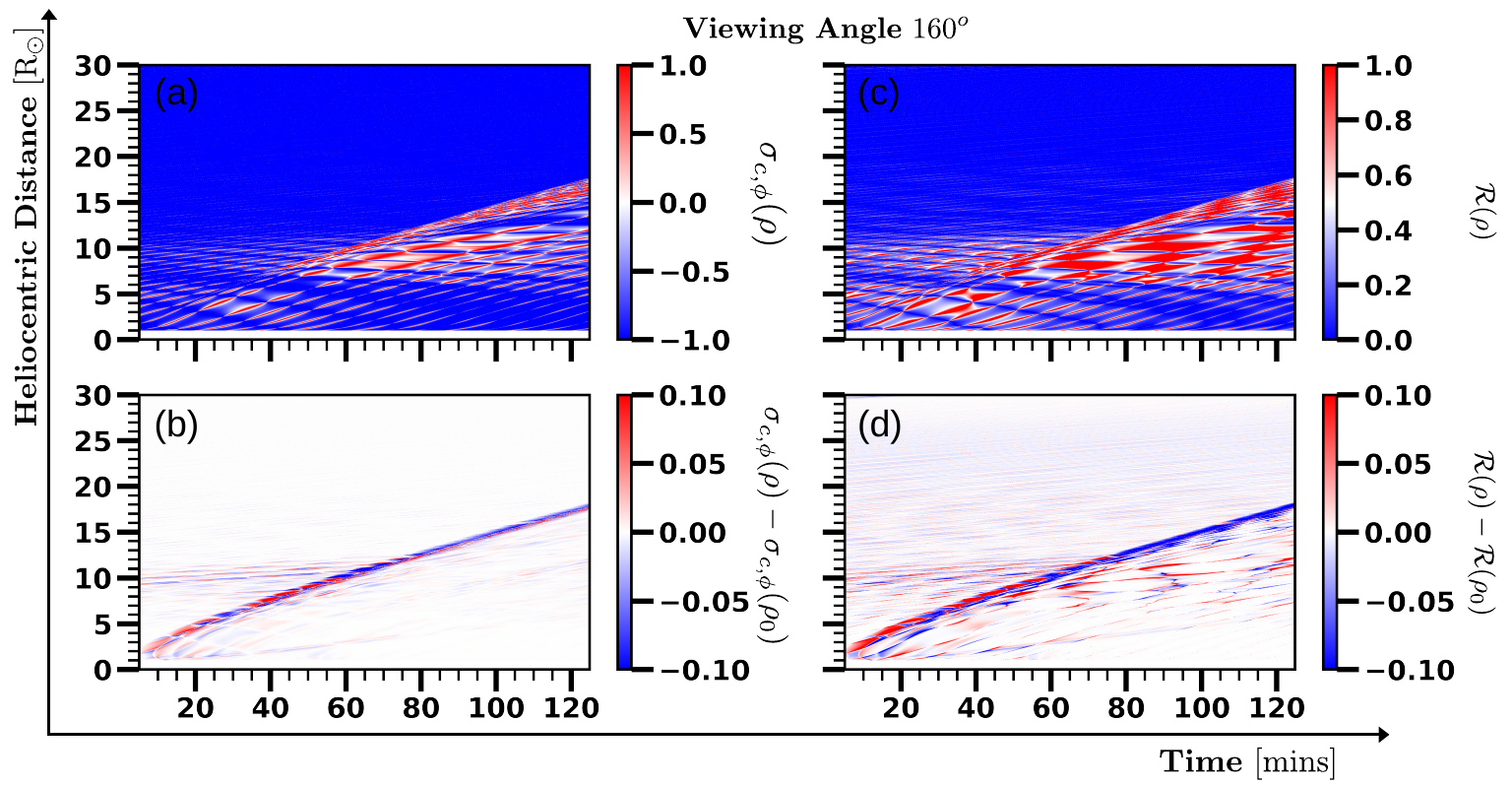}
		\caption{Effect of density fluctuations on cross helicity and the reflection coefficient. The figure presents the cross helicity (a) and the reflection coefficient (c) for the full density $\rho = \rho_0 + \delta\rho$. We evaluated the deviations in these calculations when considering only the mean density $\rho_0$ in panels (b) and (d).}
		\label{fig:cross-reflection}
	\end{figure*}
	
	In Figure~\ref{fig:cross-reflection}(a),~(b), we show the cross helicity with the full density $\rho$ and the deviation in the cross helicity calculations when assuming incompressibility using $\rho_0$, respectively. Panel (a) indicates that the plasma predominantly contains anti-sunward fluctuations ($\sigma_{c,\phi} = -1$) beyond $10~R_\odot$ in the pristine wind. There are regions containing sunward fluctuations ($\sigma_{c,\phi} = 1$) in the pristine wind at distances of less than $10~R_\odot$. The region downstream of the shock contains regions that possess all three categories: sunward, anti-sunward, and balanced fluctuations ($\sigma_{c,\phi} = 0$). Panel (b) describes the cross helicity to have a deviation of $\approx 0.1$ in the pristine wind, especially around $10~R_\odot$, where we have significant density fluctuations. 
	
	\subsection{Reflection coefficient}
	The reflection coefficient captures the fraction of $\mathbf{z}^-$ reflected to form $\mathbf{z}^+$ (or vice versa). Thus, we can define the reflection coefficient to be
	\begin{align}
		\mathcal{R} = \frac{|\mathbf{z}^+|}{|\mathbf{z}^- + \mathbf{z}^+|}.
	\end{align}
	Then, for a given $\mathcal{R}$, we can find the number of sunward waves ($|\mathbf{z}^+|$) generated due to the propagating anti-sunward waves ($|\mathbf{z}^-|$) (or vice versa). Similarly, when restricting ourselves to the $\phi$ direction,
	\begin{align}
		\mathcal{R} = \frac{|\mathbf{z}^+_{\phi}|}{|\mathbf{z}^-_{\phi} + \mathbf{z}^+_{\phi}|} = \frac{|\mathbf{z}^+_{\phi,0} - \Delta|}{|\mathbf{z}^-_{\phi,0} + \mathbf{z}^+_{\phi,0}|} 
		\label{eq:reflection-coeff}.
	\end{align}
	This implies that even in the absence of any sunward fluctuations ($\mathbf{z^+_{\phi}} = 0$), we get $\mathcal{R} > 0$ since $|\Delta| \geq 0$.
	
	Figure~\ref{fig:cross-reflection}(c) and (d) show the reflection coefficient with the full density ($\rho$) and the deviation assuming incompressibility ($\rho_0$), respectively. In the pristine solar wind, we expect predominantly anti-sunward fluctuations ($\mathcal{R} = 0$) except in locations $<10~R_\odot$, where we encounter some sunward fluctuations as well (Figure~\ref{fig:elssaser_160}(a),~(b)). Figure~\ref{fig:cross-reflection}(c) validates these conclusions with $\mathcal{R} = 0$ at $> 10~R_\odot$ but $\mathcal{R} \approx 0.5-0.8$ at $< 10~R_\odot$ in the pristine wind. We observed specks of higher $\mathcal{R}$ at all distances in the pristine wind corresponding to the density fluctuations observed in Figure~\ref{fig:results/density_waves}. The presence of density fluctuations generates a deviation of $\approx 0.1$ in $\mathcal{R}$, as seen in Figure~\ref{fig:cross-reflection}(d). The similar amplitudes of the Els\"asser variables downstream of the CME shock result in higher values of $\mathcal{R}$ along with the deviations due to the density fluctuations. If we consider the location near $10~R_\odot$, we observe an $\mathcal{R}\approx 0.5-0.7$ with a deviation of $\approx 0.05$, which corresponds to a difference of $\approx 7\%-10\%$ in the reflection coefficient. Therefore, this analysis reveals that if we transition to a solar wind containing small-amplitude compressive wave modes ($\delta\rho/\rho \leq 0.25$ in the pristine wind), then we overestimate the reflection coefficient by $\approx7\%-10\%$. We note that Alfv\'en waves are reflected by density fluctuations due to the resulting enhancements in the Alfv\'en velocity gradients~\citep{van2016heating}. Thus, the deviations in $\mathcal{R}$ discussed in this section do not refer to physical reflections of the Alfv\'en waves that would occur in such a medium. Instead, they refer to the numerical deviations introduced by the density fluctuations in the calculations in Equation~\ref{eq:reflection-coeff}. 
	
	\section{Conclusion}     \label{sec:conclusion}
	This study has described the limitations of using Els\"asser variables to analyse Alfv\'enic fluctuations interacting with a CME in the presence of small-amplitude density fluctuations. In particular, we have discussed the misinterpretations caused by the Els\"asser formalism when separating counter-propagating Alfv\'en waves in the compressive plasma. In our simulation, the compressible fluctuations in the solar wind evolved naturally through the decay of a linearly polarised Alfv\'en wave injected at the lower coronal boundary. The CME was introduced into the simulation by modelling the FR using the Grad-Shafranov equation and populating this magnetic ejecta with a non-uniform density profile in order to ensure a smooth transition to the solar wind without abrupt changes. 
	We found the solar wind plasma to be largely Alfv\'enic in nature and to exhibit strong $\mathbf{v}-\mathbf{B}$ correlations. The fluctuations in the pristine solar wind are transferred downstream of the CME shock into the sheath where the $\phi$ polarised Alfv\'en waves are scattered further in the $\theta$ direction due to density fluctuations. By confining ourselves to observing frequencies around that of the injected Alfv\'en wave, the compression of plasma at the CME shock resulted in $\delta\rho/\rho \approx 0 $ around the frequency range we investigated. This allowed us to investigate the deviations caused by the Els\"asser formalism due to the density fluctuations in the pristine wind, which has a maximum of $\delta\rho/\rho\approx 0.25$ and contrasts with the CME shock where we would have a minimal $\delta\rho/\rho$.
	
	The small-amplitude nature of the density fluctuations was validated through Figure~\ref{fig:pearsonr}, which confirms that the solar wind is dominantly  Alfv\'enic in nature. This implies that most of the wave power lies in the Alfv\'enic fluctuations. Subsequently, upon defining a parameter to quantify the difference in the Els\"asser variables with and without density fluctuations in Equation~\ref{eq:elss-eps}, we found the density fluctuations generate a maximum of $\approx 15~$km s$^{-1}$ deviations in the Els\"asser variables in a region of the pristine wind where $\delta\rho/\rho\approx 0.25$. This contributes to a difference of $\approx 4\%-10\%$ in the amplitude of anti-sunward and sunward Els\"asser variables. This deviation in the Els\"asser formalism further cascades into our interpretation of the composition of Alfv\'enic fluctuations as described in Section~\ref{sec:discussion}. We find no deviations if we only have sunward or anti-sunward fluctuations. However, in regions containing both counter-propagating Alfv\'enic fluctuations, the cross helicity calculations are deviated by $\approx 0.1$. Similarly, the reflection coefficient is overestimated by a maximum of $\approx10\%$ due to the compressible wave modes. Therefore, this study attempted to quantify the misinterpretations introduced by the Els\"asser formalism when analysing a highly Alfv\'enic solar wind where the Alfv\'enic components drive the compressive density fluctuations. This small-amplitude nature of the density fluctuations enables us to decompose the Els\"asser variables into the compressible and incompressible components, as in Equation~\ref{eq:elss-error}. These features of high Alfv\'enicity and low $\delta\rho/\rho$ are similar to the properties found in the heliospheric wind~\citep{bruno2013solar, chen2016recent}, which allows for the Els\"asser formalism-based in situ studies of solar wind plasma. While the inability of the Els\"asser formalism to exactly separate the counter-propagating Alfv\'enic waves in the presence of magnetoacoustic wavemodes is known~\citep{marsch1987ideal, magyar2019nature}, in this study, we attempted to quantise this interpretation in a more realistic simulation of a CME interacting with the solar wind.
	
	In a broader scenario, apart from the ponderomotive density fluctuations, there are also magnetosonic waves present that contribute to the driving of density fluctuations. Furthermore, the correlations between velocity and magnetic field ($\mathbf{v}-\mathbf{B}$) are not exact, resulting in regions in Figure~\ref{fig:pearsonr} where Alfvénic behaviour is absent. Additionally, the generation of density fluctuations involves multiple sources and entails non-linear interactions. In this scenario, it becomes complex to separate the Alfv\'en waves from non-Alfv\'en waves in the spatio-temporal domain~\citep{gan2022existence, fu2022nature}. However, if we further confine ourselves to specific regions that exhibit strong $\mathbf{v}-\mathbf{B}$ correlations, the non-zero $\delta\rho/\rho$ still generate quantifiable deviations in the Els\"asser formalism and their derived quantities, as we have shown. Thus, in the use of Els\"asser variables to analyse the plasma, it is imperative to note the possible amplitudes of density fluctuations and find their significance to the calculated $\mathbf{z}^\pm$ variables. Such a consideration would allow us to confidently use Els\"asser variables while being aware of the uncertainty in our calculations. For instance, in our simulation, we can mathematically find the anti-sunward waves to be $\approx 50~\mathrm{km s^{-1}}$ with a maximum deviation of $\approx 5-10~\mathrm{km s^{-1}}$. Therefore, even though we have a compressible plasma, we can still comment on the presence of some sunward Alfv\'nic fluctuations without knowing the specific means through which this reflected wave is generated. 
	
	
	\begin{acknowledgements} 
		The work has been supported by the Finnish Centre of Excellence in Research on Sustainable Space (FORESAIL; grant no. 336807). This is a project under the Academy of Finland, and this research has been supported by the European Research Council (SolMAG; grant no. 724391) as well as Academy of Finland project SWATCH (343581).
	\end{acknowledgements}
	
	\bibliographystyle{aa-note} 
	\bibliography{example}      
	
\end{document}